\documentclass[conference]{IEEEtran}
\IEEEoverridecommandlockouts
\usepackage{cite}
\usepackage{amsmath,amssymb,amsfonts}
\usepackage{mathtools}
\usepackage{algorithmic}
\usepackage{graphicx}
\usepackage{textcomp}

\usepackage{inputenc}
\usepackage{array}
\usepackage{makecell}

\usepackage{tikz}
\usepackage{pgfplots}
\usepackage{pgf}
\usepackage{wrapfig}

\usepackage[english]{babel}
\usepackage{blindtext}

\definecolor{myblue}{rgb}{0.00000,0.44700,0.74100}
\definecolor{myred}{rgb}{0.85000,0.32500,0.09800}
\definecolor{mycolor3}{rgb}{0.92900,0.69400,0.12500}
\definecolor{mypurple}{rgb}{0.49400,0.18400,0.55600}
\definecolor{mygreen}{rgb}{0.46600,0.67400,0.18800}

\makeatletter
\renewcommand\section{\@startsection{section}{1}{\z@}{1.5ex plus 1.5ex minus 0.5ex}%
{0.7ex plus 1ex minus 0ex}{\normalfont\normalsize\centering\scshape\bfseries}}%
\makeatother

\def\BibTeX{{\rm B\kern-.05em{\sc i\kern-.025em b}\kern-.08em
    T\kern-.1667em\lower.7ex\hbox{E}\kern-.125emX}}
\begin{document}

\title{\fontdimen2\font=7pt Precoding for Dual Polarization Soliton Transmission\vspace{-2mm}}
\author{\IEEEauthorblockN{A. Span\IEEEauthorrefmark{1,2}, V. Aref\IEEEauthorrefmark{2}, H. B\"ulow\IEEEauthorrefmark{1},  S. ten Brink\IEEEauthorrefmark{1}}
\IEEEauthorblockA{\IEEEauthorrefmark{1}Institute of Telecommunications, University of Stuttgart, Stuttgart, Germany}
\IEEEauthorblockA{\IEEEauthorrefmark{2}Nokia Bell Labs, Stuttgart, Germany}\vspace{-4mm}}

\maketitle

\begin{abstract}
We consider dual-polarized optical multi-soliton pulse transmission. Modulating the nonlinear spectrum, it becomes highly correlated in noisy fiber links.
We propose a simple precoding to significantly reduce the correlation, allowing disjoint detection of spectral amplitudes.
\end{abstract}

\section{Introduction}

For the nonlinear optical fiber channel, the Nonlinear Fourier Transform (NFT) maps a pulse to a nonlinear spectrum in which the pulse propagation is described by a simple linear transformation~\cite{Yousefi1}. Different ways to modulate the nonlinear spectrum have been studied and demonstrated experimentally \cite{Turitsyn}.
The evolution equations of the nonlinear spectrum are only exact in an ideal fiber where the nonlinear frequencies are preserved and their spectral amplitudes remain uncorrelated. Those properties allow to revert the transformation of each spectral amplitude in the receiver independently, known as \textit{Inverse Scattering Transform} (IST). It can also be applied in two polarizations if the optical field propagation follows the Manakov equations~\cite{gaiarin2017experimental,Goossens}. A dual-polarization $N$-th order soliton pulse is characterized in the nonlinear spectrum by $N$ 4-tuples of eigenvalues $\{\lambda_k=\omega_k+j\sigma_k\}$ and corresponding Jost (spectral) coefficients $\{a(\lambda_k),b_1(\lambda_k),b_2(\lambda_k)\}$ which can be modulated~\cite{gaiarin2017experimental}. 
In the presence of noise or other small perturbations, the IST is still practically applicable, but the eigenvalues may change (random walk fluctuations along the link) and become correlated. Accordingly, the Jost coefficients  become correlated as their transformation is related to the eigenvalues. As a consequence, the naive IST becomes less effective. 
These problems, and some more efficient detection schemes for specific cases, have been reported for single polarization~\cite{Buelow,Aref-full,gui2017alternative,Aref4}.
We propose a simple precoding scheme to transform $b_1(\lambda_k)$, $b_2(\lambda_k)$ into another pair with much less mutual correlation. The underlying idea is that $b_1(\lambda_k)$, $b_2(\lambda_k)$ undergo a similar transformation even though $\lambda_k$ is changing randomly along the link. Therefore, the differential phase of $b_1(\lambda_k)$ and $b_2(\lambda_k)$ will be well preserved along the transmission. Thus, we encode $b_1(\lambda_k)$ and $b_2(\lambda_k)$ by a common phase $\varphi_c$ and a differential phase $\varphi_d$. Then, $\varphi_d$ can be detected independently of the inverse scattering of $b_1(\lambda_k)$, $b_2(\lambda_k)$ while $\varphi_c$ must be detected via IST. 
We verify the performance of our precoding scheme in terms of reducing the estimation error via Split-Step Fourier simulation. We transmit second order solitons with QPSK modulated spectral coefficients, for a $16\mathrm{\frac{Gb}{s}}$ gross rate over $4350\mathrm{km}$ optical fiber.

\section{Transformation of Nonlinear Spectrum}

As a multi-soliton propagates along an ideal fiber, 
the spectral coefficients $b_1\left(\lambda_k\right)$, $b_2\left(\lambda_k\right)$ are transformed by \eqref{eq:b_evolution_ideal}. In the presence of noise, the eigenvalues, and accordingly, the spectral coefficients may be perturbed. However, it is not yet fully known how the transformation \eqref{eq:b_evolution_ideal} should be modified. For single-polarization solitons, a transformation is approximated using a simple Markov model~\cite{Aref4,gui2017alternative}. 
Following the arguments in \cite{Aref4}, we extend this model to dual polarization \eqref{eq:b_evolution}, where $\lambda_k(z)$ is the actual eigenvalue at distance $z$.
\begin{align}
    b_i\left(\lambda_k,L\right) & =b_i\left(\lambda_k,0\right) e^{-4j\lambda_k^2 L } 					\label{eq:b_evolution_ideal}
    \\
    b_i\left(\lambda_k,L\right) & \approx b_i\left(\lambda_k,0\right) e^{-4j\int_0^L\lambda_k^2(z) \partial z } \label{eq:b_evolution}
\end{align}

Note that instantaneous changes in eigenvalues cause some instantaneous changes in $b_i(\lambda_k(z))$, which are neglected in this simple model. In the ideal (noiseless) case, when eigenvalues do not change, this approximation becomes identical to \eqref{eq:b_evolution_ideal}. According to \eqref{eq:b_evolution}, fluctuations in the eigenvalues therefore affect the phase rotation of the spectral coefficients. This phase rotation needs to be compensated at the receiver to estimate the transmitted phases. If the information about the random walk of the eigenvalues $\lambda_k(z)$ was known, the phase rotation could be compensated almost perfectly~\cite{Aref4}. However, this information is not fully available at the receiver causing errors in the phase estimations.

\section{Precoding of the Spectral Coefficients}

\begin{figure*}
\centering
\includegraphics[scale=0.6]{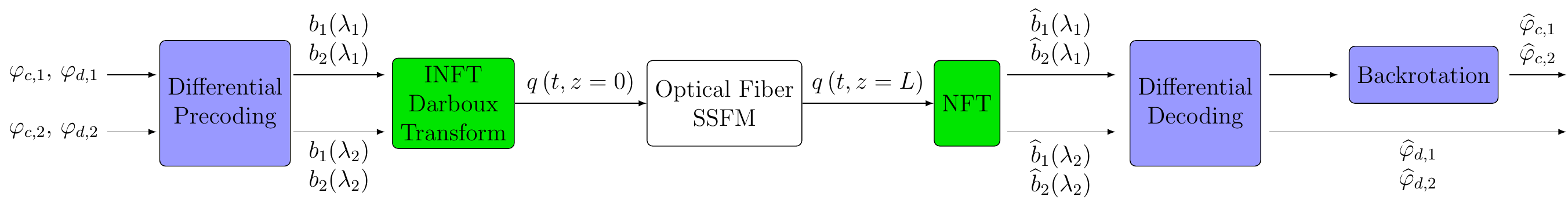}
\vspace{-2mm}
\caption{Transmission simulation setup of pol-mux 2-soliton pulses including differential precoding/decoding of $\{b_1(\lambda_k),b_2(\lambda_k)\}$, $k=1,2$. }
\label{fig:trans_setup}
\vspace{-4mm}
\end{figure*}

Although $\lambda_k(z)$ may not be fixed and fluctuate randomly,
Eq.~\eqref{eq:b_evolution} implies that
$b_1(\lambda_k,L)$ and $b_2(\lambda_k,L)$ are transformed (almost) identically. Consequently, the phase difference
between $b_1(\lambda_k,z)$ and $b_2(\lambda_k,z)$ is (almost) preserved along the link, regardless of eigenvalues fluctuations. This property suggests the following differential precoding: For each eigenvalue $\lambda_k$, both spectral coefficients $b_{1,k}\coloneqq b_1(\lambda_k)$ and $b_{2,k}\coloneqq b_2(\lambda_k)$ can be mapped bijectively to a common phase $\varphi_c$ and a differential phase $\varphi_d$ according to 
\begin{equation}\label{b_modulation}
b_{1,k} =|b_{1,k}| e^{j \varphi_{c,k}}, \quad b_{2,k} =|b_{2,k}| e^{j (\varphi_{c,k}+\varphi_{d,k})}.
\end{equation}
One can modulate the phases of $\{b_{1,k},b_{2,k}\}$, or alternatively $\{\varphi_{c,k},\varphi_{d,k}\}$. With \eqref{eq:b_evolution}, $\varphi_{d,k}$ can then be obtained from
\begin{align}\label{b1_b2_conj}
& b_{1,k}^*(L) b_{2,k}(L) \approx |b_{1,k}||b_{2,k}| e^{16\int_0^L \omega_k(z) \sigma_k(z) \partial z } e^{j\varphi_{d,k}}.
\end{align}
The estimation of $\varphi_{c,k}$ requires, however, the trace of $\lambda_k(z)$ along the link in \eqref{eq:b_evolution}. A simple estimation of the integral in \eqref{eq:b_evolution} is $\frac{L}{2}(\lambda_k^2(0)+\widehat{\lambda}^2_k(L))$, where $\lambda_k^2(0)$ is the design eigenvalue and  $\widehat{\lambda}_k(L)$ is the detected eigenvalue at the receiver.
Let $\widehat{b_i}(\widehat{\lambda}_k,L)$ denote the received spectral coefficients. Then, we estimate the differential phase $\varphi_{d,k}$ and the common phase $\varphi_{c,k}$ by \eqref{phi_d_est}. We observe $\widehat{\varphi}_{d,k}$ being independent of the eigenvalue fluctuations along the link.\vspace{-0.1mm}
\begin{align}\label{phi_d_est}
\widehat{\varphi}_{d,k}&=\arg\{\widehat{b_1}^*(\widehat{\lambda}_k,L) \widehat{b_2}(\widehat{\lambda}_k,L)\},\hspace{1 cm}		\nonumber
\\
\widehat{\varphi}_{c,k}&=\arg\{\widehat{b_1}(\widehat{\lambda}_k(L)) \exp(2j(\lambda_k^2(0)+\widehat{\lambda}_k^2(L))L)\}
\end{align}
This precoding scheme also includes two special cases: setting 
$\varphi_{d,k}=-2\varphi_{c,k}$ leads to the phase-conjugated
pair $\{b_{1,k},b_{2,k}\}$, resembling the phase-conjugated twin-wave technique in time-domain; 
$\varphi_{c,k}=0$ leads to a pilot-assisted scenario to estimate the back-rotation more precisely.

\section{Simulation Results}

\begin{figure}[h]
\vspace{-4mm}

\begin{tikzpicture}[baseline=(current axis.south)]
\begin{axis}[ticklabel style={font=\footnotesize},width=0.55\columnwidth,height=0.42\columnwidth,xmin=0,xmax=5,ymin=0,ymax=0.1,y label style={yshift=-1em,xshift=-0.5em},label style={font=\footnotesize},ylabel={Standard Deviation $(\widehat{\varphi})$ $[\mathrm{rad}]$},xlabel={distance $\ell\quad[\mathrm{km}]$},legend style={draw=none,legend pos=north west},legend style={font=\scriptsize,legend columns=1,legend image post style={xscale=0.5},fill opacity=0.1,text opacity=1},xtick={0,2,4},xticklabels={$0$,$1740$,$3480$},ytick={0,0.02,0.04,0.06,0.08,0.1},yticklabels={$0$,$2$,$4$,$6$,$8$,$10$},legend cell align=left]
    
    \addplot[forget plot,thick,mypurple,width=\linewidth/2]table {std_dev_ldv1-3.dat};
    \addplot[forget plot,thick,purple,width=\linewidth/2]table {std_dev_ldv1-4.dat};   
    \addplot[forget plot,thick,mygreen,width=\linewidth/2]table {std_dev_ldv1-5.dat};

\addlegendimage{line legend,mypurple}
\addlegendentry{$\widehat{\varphi}_{c,1}=\arg \, \widehat{b}_1(\lambda_1)$}
\addlegendimage{line legend,purple}
\addlegendentry{$\arg \, \widehat{b}_2(\lambda_1)$}
\addlegendimage{line legend,mygreen}
\addlegendentry{$\widehat{\varphi}_{d,1}$}
          
\end{axis}

\node [below=0.5cm, align=flush center,text width=1cm] at (0,3.2) {\footnotesize $\cdot 10^{-2}$};
\node [below=0.5cm, align=flush center,text width=1cm] at (0,0) {(a)};

\end{tikzpicture}
\begin{tikzpicture}[baseline=(current axis.south)]
\begin{axis}[ticklabel style={font=\footnotesize},width=0.55\columnwidth,height=0.42\columnwidth,xmin=0,xmax=5,ymin=0,ymax=4,y label style={yshift=-1em},label style={font=\footnotesize}
,xlabel={distance $\ell\quad[\mathrm{km}]$},title style={font=\small},legend style={draw=none,legend pos=north west},legend style={font=\scriptsize,legend image post style={xscale=0.5},legend columns=1,fill opacity=0.1,text opacity=1},xtick={0,2,4},xticklabels={$0$,$1740$,$3480$},ytick={0,1,2,3,4},yticklabels={$0$,$1$,$2$,$3$,$4$},legend cell align=left]

    \addplot[forget plot,thick,myblue,width=\linewidth/2]table {std_dev_ldv2-3.dat};
    \addplot[forget plot,thick,myred,width=\linewidth/2]table {std_dev_ldv2-5.dat};
    \addplot[forget plot,thick,cyan,width=\linewidth/2]table {std_dev_ldv2-4.dat};

\addlegendimage{line legend,myblue}
\addlegendentry{$\widehat{\varphi}_{c,2}=\arg \, \widehat{b}_1(\lambda_2)$}
\addlegendimage{line legend,cyan}
\addlegendentry{$\arg \, \widehat{b}_2(\lambda_2)$}
\addlegendimage{line legend,myred}
\addlegendentry{$\widehat{\varphi}_{d,2}$}

\end{axis}

\node [below=0.5cm, align=flush center,text width=1cm] at (0,3.2) {\footnotesize $\cdot 10^{-2}$};
\node [below=0.5cm, align=flush center,text width=1cm] at (0,0) {(b)};

\end{tikzpicture}


\caption{Phase estimation error as standard deviation of estimated transmitted phases along the link for (a) eigenvalue $\lambda_1$ and (b) eigenvalue $\lambda_2$}
\label{fig:sim_res}

\vspace{-2mm}
\end{figure}
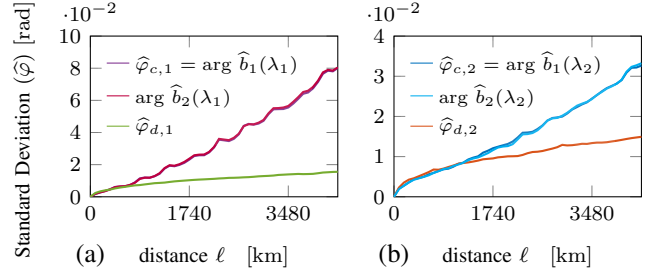

We verified the benefit of the precoding 
by transmission simulation of $2$-solitons.
The optical single-mode fiber is modeled by Manakov equations with
$\beta_2=-5.75\mathrm{\frac{ps^2}{km}}$, nonlinearity $\gamma=1.6\mathrm{\frac{1}{W \cdot km}}$ and attenuation $\alpha=0.046 \mathrm{\frac{1}{km}}$. Ideal
distributed Raman amplification was assumed with noise power spectral density \hbox{$N_{\rm ASE}=n_{sp}\alpha h\nu_s [\mathrm{\frac{W}{Hz\,km}}]$} ($n_{sp}=1.1$, $\nu_s=193.4 \mathrm{THz}$ and Planck constant $h$). We assumed an inline filtering with $50 \mathrm{GHz}$ bandwidth.
The solitons have two eigenvalues: $\lambda_1=0.5j$, $\lambda_2=1j$. The phases of $b_i(\lambda_k)$ are modulated according to \eqref{b_modulation}, where $\varphi_c$ and $\varphi_d$ are randomly chosen from a QPSK constellation.
\begin{wrapfigure}{l}{0.47\columnwidth}
\vspace{-1mm}
\hspace{-3mm}
\begin{tikzpicture}[baseline=(current axis.south)]
\begin{axis}[ticklabel style={font=\footnotesize},width=0.53\columnwidth,height=0.53\columnwidth,xmin=-1.1,xmax=1.1,ymin=-1.1,ymax=1.1,y label style={yshift=-2em},x label style={yshift=1em},label style={font=\footnotesize},ylabel=$Q$,xlabel={$I$},title style={font=\small},legend style={legend pos=north west,font=\small},xtick={},xticklabels={},ytick={},yticklabels={},legend style={font=\scriptsize,align=left},legend style={draw=none,at={(-0.28,-0.18)}},legend columns=2]

    \addplot[forget plot,only marks,myred,mark size=1pt]table {QPSK-1.dat};
    \addplot[forget plot,only marks,myblue,mark size=1pt]table {QPSK-2.dat};
    \addplot[forget plot,only marks,mygreen,mark size=1pt]table {QPSK-3.dat};
    \addplot[forget plot,only marks,mypurple,mark size=1pt]table {QPSK-4.dat};
    \draw(axis cs:0,0) circle[lightgray, radius=100];
    \draw(axis cs:0,0) circle[lightgray, radius=90];
    \draw(axis cs:0,0) circle[lightgray, radius=80];
    \draw(axis cs:0,0) circle[lightgray, radius=70];
    \draw [lightgray](0,110)--(220,110);
    \draw [lightgray] (110,0)--(110,220);

\addlegendimage{line legend,mypurple}
\addlegendentry{$\exp(j\widehat{\varphi}_{c,1})$}
\addlegendimage{line legend,mygreen}
\addlegendentry{$\exp(j\widehat{\varphi}_{d,1})$}
\addlegendimage{line legend,myblue}
\addlegendentry{$\exp(j\widehat{\varphi}_{c,2})$}
\addlegendimage{line legend,myred}
\addlegendentry{$\exp(j\widehat{\varphi}_{d,2})$}

\end{axis}


\end{tikzpicture}

\caption{Received QPSK constellation at $L=4350\mathrm{km}$ for estimated common and differential phases $\{\widehat{\varphi}_{d,k},\widehat{\varphi}_{c,k}\}$ for both eigenvalues}
\label{fig:QPSK_Rx}
\vspace{-2.5mm}
\end{wrapfigure}
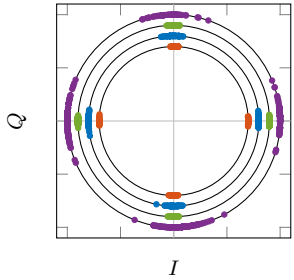
The dual-polarization soliton was generated using the Darboux transformation~\cite{Wright}. It was scaled to the physical unit and truncated with the pulse interval of $500 \mathrm{ps}$, containing more than $99.9\%$ of the pulse energy. It results in the gross rate of $16 \mathrm{\frac{Gb}{s}}$. We used the Split-step Fourier method (SSFM) for transmission simulation up to $L=4350\mathrm{km}$ fiber length. 
The average launch power was set to $-0.75 \mathrm{dBm}$ (PAPR $\approx 12 \mathrm{dB}$), resulting the received {SNR $= 27\mathrm{dB}$} at the link end. We applied Forward-backward NFT~\cite{Aref1} (extension to dual-polarization)
to the signal at different locations along the link $z$ and estimated the actual spectrum $\{\widehat{\lambda}_k(z),\widehat{b}_1(\widehat{\lambda}_k,z),\widehat{b}_2(\widehat{\lambda}_k,z)\}$. Accordingly, we estimated $\{\widehat{\varphi}_{d,k},\widehat{\varphi}_{c,k}\}$ from \eqref{phi_d_est}
and also individually estimated  $\arg \, b_1(\lambda_k)$ and $\arg \, b_2(\lambda_k)$ by IST. The simulation setup is illustrated in Fig.~\ref{fig:trans_setup} while Fig. \ref{fig:QPSK_Rx} shows the estimated phases at $L=4350$~km. We observe that both $\widehat{\varphi}_{d,k}$ were estimated with very small variations but 
$\widehat{\varphi}_{c,k}=\arg \, b_1(\lambda_k)$ is scattered with larger variations. We observe that $\arg \, b_2(\lambda_k)$ is as scattered as $\widehat{\varphi}_{c,k}$. Fig.~\ref{fig:sim_res} shows the phase variations in terms of standard deviation along the link. We observe $\widehat{\varphi}_{d,k}$ being much more reliable than $\widehat{\varphi}_{c,k}$
or $\arg \, b_2(\lambda_k)$ as the differential phase does not need the backrotation.

\section{Conclusion}

We consider the transmission of dual polarization multi-soliton pulses, phase modulated in the nonlinear spectrum.
In the presence of perturbation, e.g. ASE noise, the spectral coefficients of different eigenvalues become correlated and should be decoded together. 
We propose a differential precoding scheme that maps the correlated phases of the spectral amplitudes to another pair of phases with much less mutual correlation. These uncorrelated phases can then be recovered more reliably. 
The benefit of this precoding is illustrated by transmission simulation of second order solitons with common and differential phases of spectral coefficients being chosen from a QPSK constellation. The resulting standard deviation of the phase estimation error indicates significant improvement of the detection performance.

\end{document}